\documentclass[aps,pre,reprint,onecolumn,groupedaddress,notitlepage]{revtex4-1}

\usepackage{amsmath}
\usepackage{amsfonts}
\usepackage{amssymb,bm}
\usepackage{verbatim}
\usepackage{graphicx}
\usepackage{caption}
\usepackage{subcaption}
\usepackage{color}
\usepackage{url}

\newcommand{\beq}{\begin{equation}}
\newcommand{\eeq}{\end{equation}}
\newcommand{\vx}{\vec{x}}

\newcommand{\vur}{\vec{u}}
\newcommand{\vu}{\vec{\hat{u}}}
\newcommand{\vbr}{\vec{b}}
\newcommand{\vb}{\vec{\hat{b}}}
\newcommand{\vk}{\vec{k}}

\newcommand{\vfr}{\vec{f}}
\newcommand{\vf}{\vec{\hat{f}}}

\newcommand{\vwr}{\boldsymbol{\omega}}

\newcommand{\vaR}{\vec{a}}

\let\oldhat\hat
\renewcommand{\vec}[1]{{\bm{#1}}}
\renewcommand{\hat}[1]{{\oldhat{\bm{#1}}}}

\begin{document}


\vspace{-3em}
\begin{minipage}[l]{\textwidth}
\noindent
Postprint version of the manuscript published in Phys.~Rev.~Fluids 
{\bf 2}, 114604 (2017). \\
\end{minipage}
\vspace{1em}

\title{Comparison of forcing functions in magnetohydrodynamics}


\author{Mairi E. McKay$^1$}
\email[]{mairi.mckay@ed.ac.uk}
\author{Moritz Linkmann$^2$}
\author{Daniel Clark$^1$, Adam A. Chalupa$^1$}
\author{Arjun Berera$^1$}
\email[]{ab@ph.ed.ac.uk}
\affiliation{$^1$ The School of Physics and Astronomy, The University of Edinburgh, Edinburgh, United Kingdom}
\affiliation{$^2$ Department of Physics \& INFN, University of Rome Tor Vergata, Via della Ricerca Scientifica 1, 00133 Rome, Italy}


\date{\today}

\begin{abstract}
Results are presented of direct numerical simulations of incompressible, homogeneous magnetohydrodynamic turbulence 
without a mean magnetic field, subject to different mechanical forcing functions 
commonly used in the literature. 
Specifically, the forces are negative damping 
(which uses the large-scale velocity field as a forcing function), 
a nonhelical random force, and a nonhelical static sinusoidal force 
(analogous to helical ABC forcing). 
The time evolution of the three ideal invariants 
(energy, magnetic helicity and cross helicity), the time-averaged energy spectra, 
the energy ratios and the dissipation ratios are examined. 
All three forcing functions produce qualitatively similar steady states 
with regards to the time evolution of the energy and magnetic helicity.
However, differences in the cross helicity evolution are observed, 
particularly in the case of the static sinusoidal method of energy injection.
Indeed, an ensemble of sinusoidally-forced simulations with identical parameters 
shows significant variations in the cross helicity over long time periods, 
casting some doubt on the validity of the principle of ergodicity in systems in which the injection of helicity cannot be controlled. 
Cross helicity can unexpectedly enter the system through the forcing function and must be carefully monitored.
\end{abstract}


\maketitle


\section{Introduction}
Turbulence is a diverse and complex phenomenon that has
been a source of interest for 100 years, if not more,
and from many different disciplines.
The Navier-Stokes equations that describe the turbulent flow 
of a nonconducting fluid
are well-known but their nonlinear nature prohibits a complete understanding of them.
To describe the behavior of conducting fluids, they are combined with
Maxwell's equations of electromagnetism
to form the magnetohydrodynamic equations.
These equations can be applied to many astrophysical and geophysical flows, 
and they also have industrial applications
\cite{Biskamp2003,Davidson2001,Frisch1995,Verma2004}.

The advent of high performance computing has caused a rapid growth 
in the study of turbulence.
Direct numerical simulations (DNS) are a computationally expensive tool
that allows us to follow the exact evolution of a turbulent flow
without introducing any modeling.
However, it is often useful to inject energy into a turbulent system
to compensate for the energy lost through dissipation,
and in these forced simulations, 
a decision has to be made about the method of energy injection.
Once a balance is achieved between the energy lost and 
the energy injected, the system's statistical properties can be studied.
Homogeneous turbulence, which allows fundamental aspects of turbulence to be 
studied without concern for additional effects from, e.g., boundary conditions,
is often simulated. Strictly speaking, an infinite
domain is required for homogeneous turbulence but
in computational studies this is replaced by periodic boundary conditions.

A wide range of approaches to forcing homogeneous turbulence simulations has
been used over the years.
Most often these involve injecting energy into the smallest wavenumbers,
with or without introducing some random component, see e.g. Refs. \cite{Eswaran1988,Alvelius1999,Zeren2010}.
Different forcing methods have different advantages.
A deterministic force could be seen as more physical
but stochastic forces may provide better control over energy and helicity input.
Another example is that a deterministic forcing may produce
fluctuations on larger time scales
than a stochastic forcing and so it may require a longer run-time to 
obtain converged statistics, despite being more efficient
computationally \cite{Zeren2010}.

At the core of the study of homogeneous isotropic turbulence
are the contributions of Richardson, Kolmogorov and Obukhov,
which include the energy cascade (the notion that energy is transferred
from large scales to progressively smaller scales where dissipation dominates),
the self-similar scaling of structure functions
and the famous `five-thirds' power law for the energy spectrum, $E(k)\propto k^{-5/3}$.
These features give rise to the concept of universality in nonconducting fluids,
which is the idea that the small-scale evolution of a turbulent system
is independent of the large-scale features,
such as geometry or the method of energy injection.
The concept of universality has been supported by many experiments, numerical simulations
and theoretical arguments since then \cite{McComb2014,Batchelor1953,Frisch1995,Pope2000}.

Universality in magnetohydrodynamics (MHD) is questionable;
highlighted especially by the unusual presence of an inverse cascade 
of magnetic helicity under certain conditions,
which causes some energy to be transferred to the large scales \cite{Leorat1975,Pouquet1976,Alexakis2005a,Alexakis2007}. 
The three ideal invariants in MHD are the total energy,
magnetic helicity and cross helicity.
The latter two quantities represent the knottedness of 
magnetic field lines
and the alignment of the velocity and magnetic fields \cite{Moffat1969}.
It is known that these parameters can alter 
aspects of MHD systems such as the (dimensionless) dissipation rate \cite{Linkmann2015,Linkmann2017,Dallas2013}.
This comes as a result of selective decay processes,
that is, the tendency for the fields to self-organize
into a force-free state in which the magnetic helicity is maximal,
or an Alfv\'enic state in which the velocity and magnetic fields are fully aligned \cite{Biskamp2003,Dobrowolny1980}.
The alignment of the fields weakens the turbulence,
slowing down the decay of energy in unforced systems \cite{Biskamp2003,Stribling1991}.
Furthermore, the presence or absence of a background
magnetic field affects the turbulent dynamics,
and this is reflected in the scaling of the energy spectrum.
In forced MHD with a strong background magnetic field, the 
theory of dynamic alignment predicts 
the field-perpendicular energy scaling
$E(k_\perp)\propto k^{-3/2}$ \cite{Boldyrev2006}.
This comes as a result of the tendency for the
 magnetic and velocity fluctuations to become aligned,
with a more pronounced effect at smaller scales.
However, dynamic alignment may also occur at small scales 
in the absence of a guide field, 
since large-scale magnetic-field fluctuations can play a similar role 
for small-scale fluctuations as the guide field \cite{Boldyrev2006}.

Interestingly, it was found that self-organised
 states can be obtained when both fields are forced in a nonhelical way, depending on 
the time-correlation of the forcing \cite{Dallas2015}.
The less often the phase was randomised, 
the more the cross helicity and magnetic helicity would build up, leading to self-organisation.
The Archontis dynamo, in which the velocity field is forced
in a nonhelical way, is also known to introduce large values of cross helicity.
Furthermore, in the case in which the magnetic and kinetic Reynolds numbers Rm and Re
are equal,
the Archontis dynamo saturates with the magnetic and kinetic energy almost in equipartition
\cite{Archontis2007,Dorch2004,Gilbert2011,Cameron2006}.
It is important to be sure that effects such as these are independent of the 
specific implementation of the forcing, especially since MHD has many cosmological,
astrophysical and industrial applications and relies heavily on results from numerical simulations.
The importance of understanding how independent
the turbulence properties are of the forcing function has been
underpinned by the above discussion.

In this paper, we present a systematic study that examines the effect of
different forcing functions in MHD.  We investigate the evolution of
homogeneous, incompressible MHD turbulence without a mean magnetic field,
subject to three different types of mechanical forcing functions that aim to
represent the range of forcing methods used in the literature.  Specifically,
we use (i) a mechanical force that uses the large-scale velocity field as a
forcing function, (ii) a nonhelical random force defined by using time-varying
helical basis vectors, and (iii) a nonhelical static sinusoidal force.  In Sec.
\ref{NM}  we give details of our simulations, including the three forcing
routines.  In Sec. \ref{R} we examine the time evolution of the three ideal
invariants (energy, magnetic helicity and cross helicity), the time-averaged
energy and cross helicity spectra, the energy ratios and the dissipation
ratios.  As we will show, the magnetic helicity remains close to zero in all
cases but the sinusoidal method of energy injection has a tendency to introduce
cross helicity into the system.  Indeed, our results for sinusoidally-forced
simulations with identical parameters and different initial conditions show
large variations in the normalised cross helicity over long time periods.  We
discuss these results in Sec. \ref{C} and draw some conclusions.

\section{Numerical method\label{NM}}
\begin{table*}[]
\centering
\begin{tabular}{cccccccccccc}
Run ID & Type & $N$  & $\nu$    & Re   & $\text{Re}_\lambda$ & $k_{max}/k_\eta$ & $k_{max}/k_\nu$ & $\rho_b$ & $\rho_u$ & $\rho_c$ & $\sigma_c$ \\ \hline
AHFa   & AHF  & 1024 & 0.000625 & 1085 & 218                 & 2.49             & 3.13            & -0.00011 & 0.0014   & -0.012   & -0.010     \\
NDa    & ND   & 1024 & 0.000625 & 1293 & 233                 & 2.60             & 3.26            & 0.0036   & 0.040    & 0.063    & 0.057      \\
SFa    & SF   & 1024 & 0.000625 & 1524 & 272                 & 2.61             & 3.27            & 0.016    & -0.0034  & 0.10     & 0.084      \\ \hline
AHFb   & AHF  & 512  & 0.0008   & 742  & 180                 & 1.50             & 1.88            & 0.0077   & 0.00063  & 0.026    & 0.023      \\
NDb    & ND   & 512  & 0.0008   & 994  & 213                 & 1.56             & 1.96            & -0.0043  & 0.045    & -0.035   & -0.030     \\
SFb    & SF   & 512  & 0.0008   & 1190 & 236                 & 1.55             & 1.94            & -0.0046  & 0.0027   & -0.16    & -0.13      \\ \hline
AHFc   & AHF  & 512  & 0.001    & 609  & 162                 & 1.77             & 2.20            & -0.0038  & -0.0072  & 0.012    & 0.010      \\
NDc    & ND   & 512  & 0.001    & 803  & 179                 & 1.85             & 2.28            & -0.029   & -0.00015 & 0.22     & 0.21       \\
SFc    & SF   & 512  & 0.001    & 940  & 207                 & 1.85             & 2.31            & 0.0022   & 0.00076  & -0.012   & -0.010     \\ \hline
AHFd   & AHF  & 256  & 0.0015   & 374  & 125                 & 1.25             & 1.51            & -0.0039  & 0.0037   & -0.0084  & -0.0069    \\
NDd    & ND   & 256  & 0.0015   & 535  & 152                 & 1.25             & 1.53            & -0.0046  & 0.044    & -0.093   & -0.080     \\
SFd    & SF   & 256  & 0.0015   & 618  & 164                 & 1.25             & 1.55            & 0.011    & -0.0046  & -0.029   & -0.025     \\ \hline
AHFe   & AHF  & 256  & 0.002    & 308  & 113                 & 1.55             & 1.82            & 0.0069   & -0.0098  & -0.0062  & -0.0049    \\
NDe    & ND   & 256  & 0.002    & 410  & 134                 & 1.56             & 1.88            & -0.013   & 0.059    & 0.0010   & 0.00081    \\
SFe    & SF   & 256  & 0.002    & 541  & 161                 & 1.45             & 1.76            & -0.0037  & 0.0015   & 0.013    & 0.010      \\ \hline
AHFf   & AHF  & 128  & 0.005    & 127  & 62                  & 1.61             & 1.61            & -0.0044  & -0.0019  & -0.0088  & -0.0060    \\
NDf    & ND   & 128  & 0.005    & 171  & 76                  & 1.58             & 1.72            & 0.0036   & -0.041   & 0.024    & 0.018      \\
SFf    & SF   & 128  & 0.005    & 211  & 89                  & 1.46             & 1.65            & 0.021    & -0.014   & 0.10     & 0.080     
\end{tabular}
\caption{Table of basic parameters including the forcing type, number of grid points $N^3$, viscosity $\nu$, integral-scale and Taylor-scale
         Reynolds numbers Re and Re$_\lambda$ respectively,
         resolution $k_{max}/k_\eta$ and $k_{max}/k_\nu$ 
         defined with respect to the magnetic and velocity fields,
	 relative magnetic helicity $\rho_b$,
	 relative kinetic helicity $\rho_u$,
 	 and two definitions of relative cross helicity $\rho_c$ and $\sigma_c$ (Eqs.~\eqref{eq:ch1} and \eqref{eq:ch2}).
	 The values were time-averaged over the duration of the steady state.
         AHF denotes adjustable helicity forcing,
         ND denotes negative damping, and SF denotes sinusoidal forcing. }
\label{tab:param}
\end{table*}

We performed direct numerical simulations of the incompressible MHD equations
\begin{align}
  &\partial_t \vur = -\nabla P - (\vur\cdot\nabla)\vur + (\nabla \times \vbr) \times \vbr + \nu \nabla^2 \vur + \vfr\\
  &\partial_t \vbr = \nabla \times (\vur \times \vbr) +\eta\nabla^2\vbr \\
  &\nabla\cdot\vur=0\text{, }  \nabla\cdot\vbr=0 \ ,
\end{align}
where $\vur$ is the velocity field, $\vbr$ is the magnetic field in Alfv\'{e}n units, 
$P$ is the pressure, $\nu$ is the kinematic viscosity, $\eta$ is the magnetic diffusivity, and 
$\vfr$ is an external force that we will define later.
The density was set to unity. We use a caret to denote the Fourier transform of a field.
We solved these equations numerically using a pseudospectral, fully-dealiased code 
(see \cite{Yoffe2012,Linkmann2016} for details)
on a three-dimensional periodic domain with lattice sizes varying from $128^3$ to $1024^3$ points
(see Tab. \ref{tab:param}). 
The initial fields were random Gaussian with magnetic and kinetic energy spectra of the form
$E_{b,u}(k,t=0)=Ak^4\exp(−k^2/(2k_0)^2)$,
where A is a positive real number and $k_0$ is the peak of the spectrum,
which we set to 5 in all cases.
There was no imposed magnetic guide field.
The simulations were all spatially well-resolved, with the maximum
resolved wavenumber always at least 1.25 times greater than the dissipation wavenumbers $k_{\nu}=(\epsilon_u/\nu^3)^{1/4}$
and  $k_{\eta}=(\epsilon_b/\eta^3)^{1/4}$ associated with the Kolmogorov microscales,
where $\epsilon_u$ and $\epsilon_b$ are the kinetic and magnetic dissipation rates.
For simplicity we limited our study to the case where the magnetic Prandtl number, Pm$=$Rm$/$Re$=\nu/\eta$, is one.

We examined the variations of the relative magnetic helicity and cross helicity.
The relative magnetic helicity is defined as
\begin{equation}
\rho_b=\langle \vbr\cdot\vaR\rangle/(\langle|\vbr|^2\rangle\langle|\vaR|^2\rangle)^{1/2}
\end{equation}
where $\vec{a}$ is the magnetic vector potential, $\vbr=\nabla\times\vec{a}$,
and the angular brackets denote a spatial average.

The relative cross helicity can be quantified in two ways:
\begin{align}
\rho_c&=\langle \vur\cdot\vbr\rangle/(\langle|\vur|^2\rangle\langle|\vbr|^2\rangle)^{1/2} \label{eq:ch1} \ \text{and}\\
\sigma_c&=2\langle \vur\cdot\vbr\rangle/(\langle |\vur|^2\rangle+\langle|\vbr|^2 \rangle) \label{eq:ch2} \ .
\end{align}
The former is a measurement of the alignment between the fields,
whereas the latter is the ratio of two ideal invariants, namely
the cross helicity and the total energy.
Together they obey the inequality $|\sigma_c|\leq|\rho_c|\leq1$
\cite{Grappin1983,Pouquet1988}. Both definitions have their merits.
Their differences and similarities are explored in Sec. \ref{CH}.
We focus our attention mostly on the larger of the two, $\rho_c$,
which is more sensitive to the forcing function.

The relative kinetic helicity is defined as
\begin{equation}
\rho_u=\langle \vur\cdot\vwr\rangle/(\langle|\vur|^2\rangle\langle|\vwr|^2\rangle)^{1/2} \ ,
\end{equation}
where $\vwr=\nabla\times\vur$ is the vorticity.

Three types of forcing function were used: negative damping (ND), adjustable helicity forcing (AHF)
and sinusoidal forcing (SF) which are defined as follows:

\subsection{Negative damping}
The negative damping function uses the large-scale velocity field as a forcing function.
It was first developed as a way to avoid introducing
further randomness into an already random system \cite{Machiels1997}
and it is commonly used in hydrodynamic simulations \cite{Jimenez1993,Kaneda2003,Kaneda2006,McComb2015,Linkmann2015a}.
The function is
\begin{align}
\label{eq:ND}
\vf(\vk,t)  = 
\begin{cases}
  \frac{\epsilon_i \vu(\vk,t)}{2E_{u, k_f}(t)}, & \text{if } 1\leq |\vk|\leq k_f \\
  0             & \text{otherwise,}
\end{cases}
\end{align}
where $E_{u, k_f}(t)=\int_1^{k_f} E_u(k,t) dk$
is the kinetic energy contained in the forcing range $[1,k_f]$
and $\epsilon_i$ is an adjustable parameter.
The rate of energy injection is
$\langle\vur\cdot\vfr\rangle = \int d\vec{k} \ \vu(\vec{k}) \cdot \vf(-\vec{k}) =\epsilon_i$
which will be equal to the mean total
dissipation rate $\epsilon=\epsilon_b+\epsilon_u$ 
during the steady state.
We chose our fields' initial conditions
to have negligible kinetic, magnetic and cross helicity
and therefore one might expect the fields to remain nonhelical
throughout their evolution, 
although the actual helicity injection cannot be controlled.
The variation of cross helicity in MHD subject to negative damping was explored to some extent in \cite{Sahoo2011}.
The forcing type has been well-used, but nevertheless, it was recently found that,
in hydrodynamics, at low Reynolds numbers,
negative damping can induce self-ordering effects 
due to poor control of kinetic helicity injection \cite{McComb2015, Linkmann2015a}.

\subsection{Adjustable helicity forcing\label{AHF}}
The second type of forcing considered uses a helical basis
composed of eigenvectors of the curl operator:
\begin{align}
\label{eq:AHF}
\vf(\vk,t)  = 
\begin{cases}
 A(\vk)\vec{e}_1(\vk,t)+B(\vk)\vec{e}_2(\vk,t), & \text{if } 1\leq |\vk|\leq k_f \\
  0             & \text{otherwise,}
\end{cases}
\end{align}
where $\vec{e}_1\cdot\vec{e}_2^*=\vec{e}_1\cdot\vk=\vec{e}_2\cdot\vk=0$
and $\vec{e}_1$ and $\vec{e}_2$ are unit vectors that statisfy
 $i\vk\times\vec{e}_1=k\vec{e}_1$ and $i\vk\times\vec{e}_2=-k\vec{e}_2$.
At each forcing time step, for every vector $\vk$ with magnitude $1 \leq |\vk| \leq k_f$, a corresponding
random perpendicular unit vector is generated
and used to construct the helical basis;
thus the basis is changed every time the forcing function is called.
$A(\vk)$ and $B(\vk)$ are complex parameters that 
can be adjusted to control the helicity of the forcing \cite{Brandenburg2001}.
In our simulations we set the kinetic helicity 
to zero, so the forcing was explicitly nonhelical. 
This type of forcing has been widely used \cite{Malapaka2013,Muller2012,Biferale2012,Brandenburg2001,Brandenburg2014}.

\subsection{Sinusoidal forcing\label{SF}}
The sinusoidal forcing we used is deterministic and nonhelical, implemented in real space:
\begin{align}
\vfr(\vx) &= C \sum_{k=1}^{k_f}
  \begin{pmatrix}
    \sin(kz)+\sin(ky) \\
    \sin(kx)+\sin(kz) \\
    \sin(ky)+\sin(kx)
  \end{pmatrix} \ ,
\end{align}
where $C$ is an adjustable constant.
This forcing type was used in Ref. \cite{Dallas2015}. 
It is a nonhelical analog of the well-known
ABC forcing, which is fully helical \cite{Galanti1992,Galloway2012,Mininni2007,Childress1970,Galloway1984},
and it is similar to the Archontis dynamo, 
$\vfr(\vx)=(\sin(z),\sin(x),\sin(y))$ \cite{Cameron2006}.

\section{Results\label{R}}
\subsection{Energy evolution}
\begin{figure}
  \centering
  \includegraphics[width=0.7\linewidth]{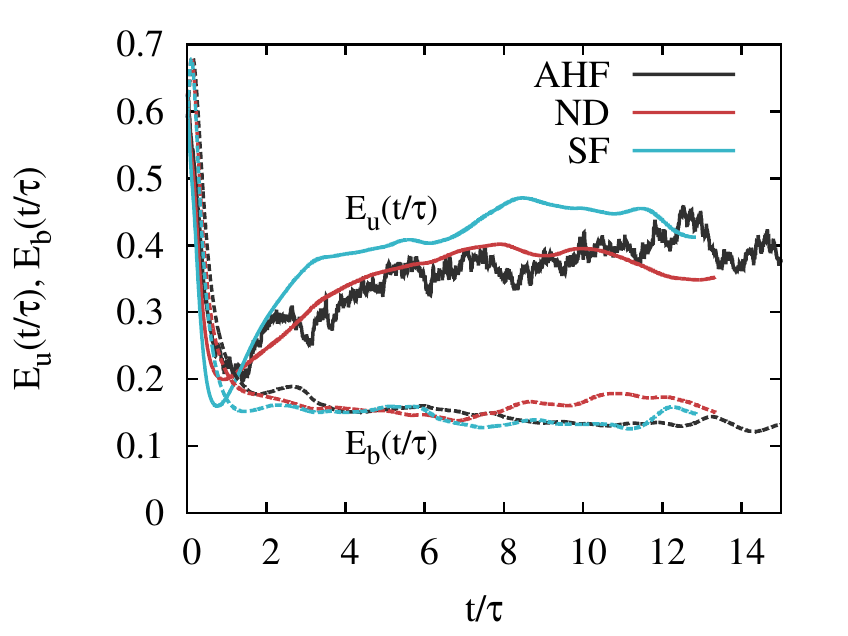}
  \caption{Evolution of kinetic energy (solid lines) and magnetic energy (dashed lines)
           for runs AHFa, NDa and SFa. 
           $\tau$ is the steady state large eddy turnover time. (Colour online.)}
  \label{fig:E}
\end{figure}
We will describe our results using the
$\nu=0.000625$ (AHFa, NDa, and SFa) simulations to demonstrate points,
since they attained the highest Reynolds numbers
in our tests (see Tab. \ref{tab:param}).
Using these simulations, which are representative of
the other simulations we ran, we will highlight features of the 
three forcing functions.

Our analysis focuses on various properties of the forced systems
while in a statistically steady state.
This allows us to look at time-averaged samples of data taken during that period.
Figure \ref{fig:E} shows the time evolution of the kinetic and magnetic energies
corresponding to runs AHFa, NDa and SFa. 
An initial transition period precedes
the fully-developed 
statistically steady turbulent state, where the energy injected
equals the energy dissipated.
We began taking measurements after
the transient initial behavior had passed and
both the kinetic and magnetic energies were fluctuating around a constant value.
The AHF kinetic energy evolution, as seen in Fig. \ref{fig:E}, is more erratic than the 
other two forcing types. 
This is due to the random nature of the forcing function,
as described in Sec. \ref{AHF}, which causes rapid changes in 
the amount of energy injected.
The time scale was normalized by the steady state 
large eddy turnover time $\tau=u_{rms}/L$, 
where $u_{rms}$ is the root-mean-square velocity,
 $L=3\pi\int_0^\infty k^{-1}E_u(k)dk/(4\int_0^\infty E_u(k)dk)$ is 
the integral scale and $E_u(k)$ is the steady state kinetic energy spectrum.
In an isotropic system, $u_{rms}^2=\langle u_i^2\rangle$
for any direction $i$, so the total kinetic energy $E_u=3u_{rms}^2/2$.
All simulations lasted for 100 units
of simulation time, corresponding to approximately 30 to 40 
large eddy turnover times, except the simulations run on $1024^3$ points,
which ran for about 40 units of simulation time. The AHF runs generally had a slightly
smaller value of $\tau$,
meaning that the injected energy was transferred to the smaller scales
at a faster rate.

We will use both the integral scale Reynolds number 
$\text{Re}=u_{rms}L/\nu$ and the Taylor Reynolds number
$\text{Re}_\lambda=u_{rms}\lambda/\nu$
as metrics to measure the turbulence,
since the integral scale Reynolds number is associated 
with the forcing scales
while the Taylor Reynolds number characterizes 
the turbulence at intermediate scales.
Here $\lambda=(15\nu/\epsilon_u)^{1/2}u_{rms}$ is the Taylor microscale.

  \begin{figure}
    \centering
    \includegraphics[width=0.9\linewidth]{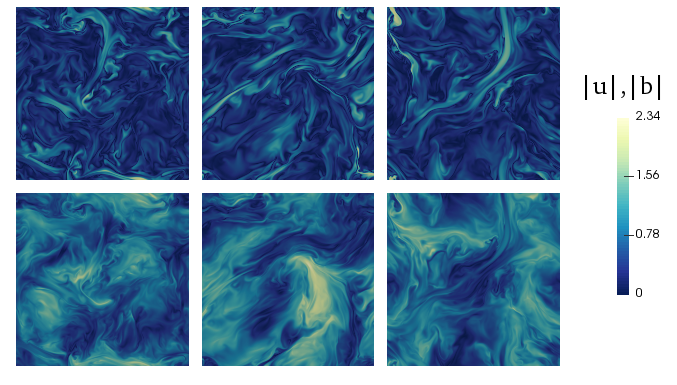}\\
    \vspace{-1em}
\hspace{-3.0cm}  AHFa \hspace{3.5cm} NDa \hspace{3.5cm} SFa
   \caption{Visualisation of a two-dimensional slice of the magnitudes of the 
             magnetic (top) and velocity (bottom) fields
             in the AHFa (left), NDa (middle) and SFa (right) cases. (Colour online.)}
    \label{fig:vis}
  \end{figure}

Two-dimensional slices of the magnitudes of the fields $|\vur|$ and $|\vbr|$ 
from the AHFa, NDa and SFa simulations
at a point in time during the steady state
are shown in Fig. \ref{fig:vis}.
These slices are representative of the general structure of the fields
throughout the steady state time frame. 
The time-averaged Reynolds numbers are moderately separated: 
$\text{Re}=1085,1293$ and $1524$ respectively,
but the fields do not differ greatly 
and exhibit the same level of detail in the small scales.
This is to be expected as they have similar dissipation 
wavenumbers $k_\eta$ and $k_\nu$ (see Tab. \ref{tab:param}).
These visualisations demonstrate that,
although the forces have very different functional forms,
the physical appearance of the fields is similar.
Overall we see that all three forces
are capable of producing physically-alike steady state behavior 
in the same time frame.

\subsection{Definitions of relative cross helicity\label{CH}}
\begin{figure}
  \begin{subfigure}{0.5\linewidth}
    \centering
    \includegraphics[width=0.9\linewidth]{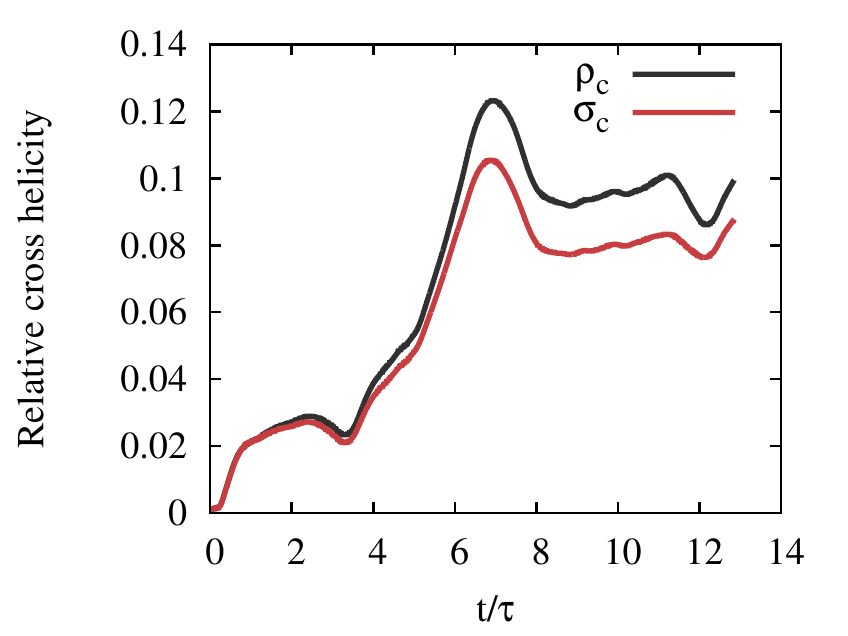}
    \caption{\label{fig:ch_comp}}
  \end{subfigure}
  \begin{subfigure}{0.49\linewidth}
    \centering
    \includegraphics[width=0.9\linewidth]{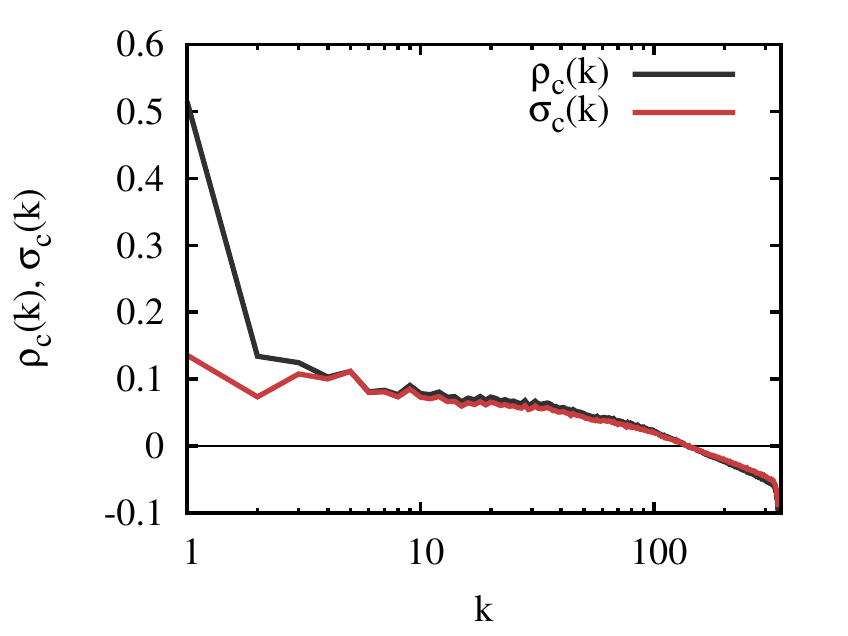}
    \caption{\label{fig:spectra_ch_comp}}
  \end{subfigure}
\caption{(a) Time evolution and (b) spectra of the two definitions of 
         relative cross helicity [Eqs.~\eqref{eq:ch1} and \eqref{eq:ch2}] for run SFa. 
         Note the logarithmic scale on the $x$-axis in (b).
         (Colour online.) \label{fig:ch_comp_main}}
\end{figure}

In Sec. \ref{NM} two ways of defining the relative cross helicity were introduced.
The time evolution of these quantities and their time-averaged spectra 
from run SFa are shown in Fig. \ref{fig:ch_comp_main}.
 
As can be seen in the figure, the
time evolution of the two metrics follow a similar trajectory, with 
$\sigma_c$ remaining close to, but less than, $\rho_c$. 
The largest separation between $\rho_c$ and $\sigma_c$
 occurs during a period when the difference between the kinetic and
magnetic energies was greatest;
this effect is apparent from the definitions of $\rho_c$ and $\sigma_c$
given in Eqs.~\eqref{eq:ch1} and \eqref{eq:ch2}.
When the kinetic and magnetic energies are in equipartition the two quantities coincide.

Figure \ref{fig:spectra_ch_comp} shows the time-averaged 
cross helicity spectra corresponding to the two definitions
in Eqs.~\eqref{eq:ch1} and \eqref{eq:ch2}
\begin{align}
\rho_c(k) &= \frac{{H_c(k)}}{2E_b(k)^{1/2}E_u(k)^{1/2}} \ , \\
\sigma_c(k) &= \frac{{H_c(k)}}{E_b(k)+E_u(k)} \ ,
\end{align}
where $H_c(k) = \int_{|\vec{k}|=k} d \vec{k} \ \hat{u}(\vec{k}) \cdot \hat{b}(-\vec{k})$ 
is the cross helicity spectrum.
We see that $\rho_c(k)$, which is normalised by 
$2E_b(k)^{1/2}E_u(k)^{1/2}$,
is more sensitive to
the ratio of 
kinetic and magnetic energy than $\sigma_c(k)$ is.
Because of this, at the forcing scale,
where the kinetic energy is much greater than the magnetic energy,
the denominator becomes small for $\rho_c(k)$ and not for $\sigma_c(k)$.
Thus $\rho_c(k)$ has a large peak 
that is absent from the $\sigma_c(k)$ spectrum.
The difference in the low-$k$ relative cross helicity spectra
occurs consistently across simulations with different forcing functions.
This effect, however, could cause data to be misinterpreted,
especially between separate groups of researchers.

\subsection{Comparison of energy and cross helicity spectra}
\begin{figure}
  \begin{subfigure}{0.5\linewidth}
    \centering
    \includegraphics[width=0.9\linewidth]{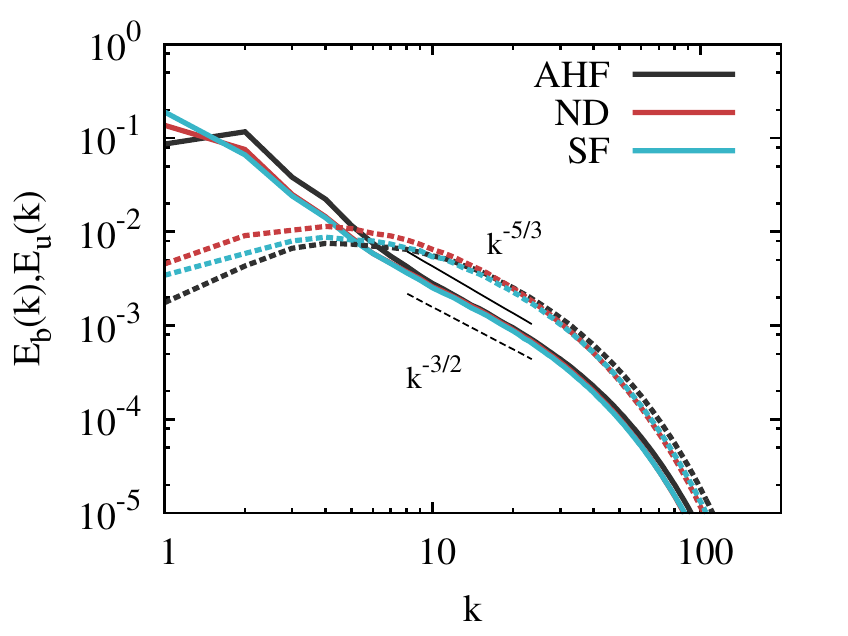}
    \caption{\label{fig:spectra}}
  \end{subfigure}
  \begin{subfigure}{0.49\linewidth}
    \centering
    \includegraphics[width=0.9\linewidth]{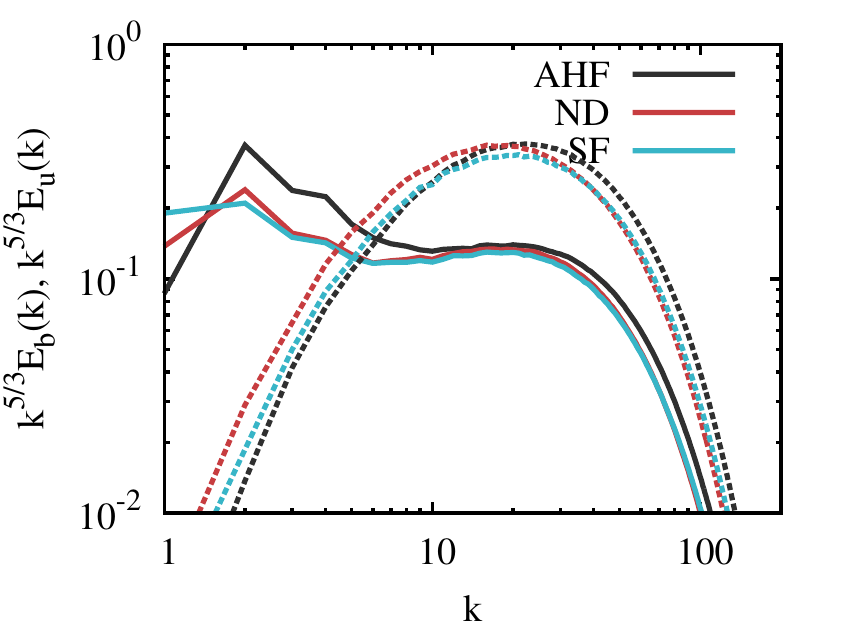}
    \caption{\label{fig:comp_spectra}}
  \end{subfigure}
\caption{(a) Kinetic and magnetic energy spectra (solid and dashed lines respectively)
         and (b) compensated kinetic and magnetic energy spectra
         for runs AHFa, NDa and SFa. (Colour online.)}
\end{figure}

\begin{figure}
  \centering
  \includegraphics[width=0.7\linewidth]{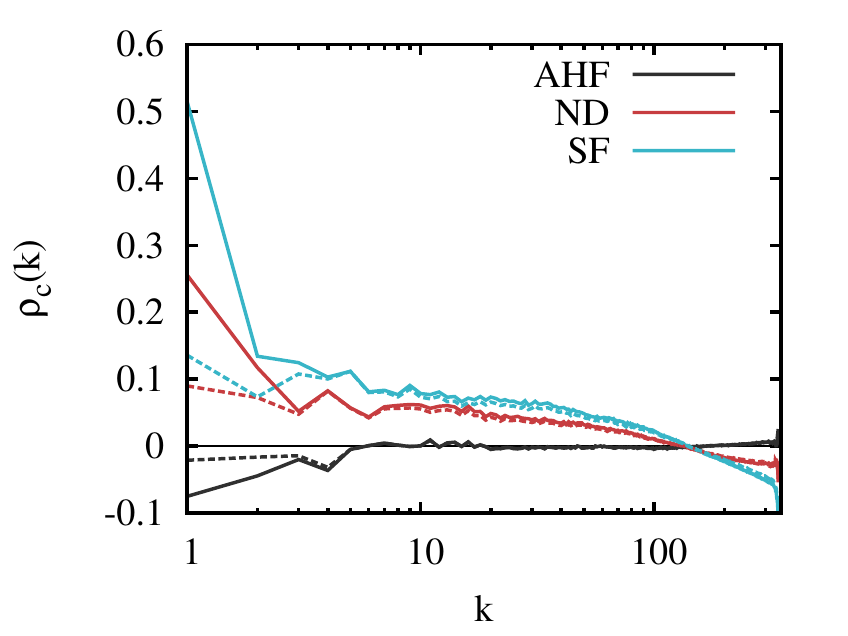}
  \caption{Relative cross helicity spectra for runs AHFa, NDa and SFa. 
           Note the logarithmic scale on the $x$-axis.
           (Colour online.)}
  \label{fig:ch_spectra}
\end{figure}

Having identified the onset of the statistically steady state, 
we can examine the time-averaged energy spectra [Fig.~\ref{fig:spectra}].
The spectra coincide in a small inertial subrange
but spread out slightly at the large and small scales.
All of our simulations had the same low-$k$ behavior,
with the ND and SF runs having a peak at $k=1$ and
the AHF types peaking at $k=2$.

Both
$k^{-5/3}$ and $k^{-3/2}$ scalings have been indicated in Fig. \ref{fig:spectra}.
The compensated kinetic and magnetic energy spectra are shown in Fig.~\ref{fig:comp_spectra}.
As can be seen in these figures, the power-law range of the 
kinetic energy spectrum is too short to distinguish 
between the two scalings.
This highlights that we still have to exercise caution in making measurements of this kind,
and that larger simulations with a more obvious inertial range are still needed.
The inertial range for the magnetic energy spectrum is even less clear. 
It is plausible that in order to see a clearer scaling
we would need to inject energy directly into the magnetic field;
this is also indicated by the results of Ref. 
\cite{Alexakis2013}
in which the ratio of magnetic to mechanical forcing was varied
in generally nonhelical simulations.
However, our results focus only on mechanical forcing, as 
magnetic forcing is known to induce other effects 
such as large-scale self-organisation \cite{Dallas2015}.
The runs with smaller Reynolds numbers have steeper energy spectra,
presumably because of the enduring problem of the lack of separation between forcing scales
and dissipation scales in direct numerical simulations.
The steeper slope in the lower range of run AHFa compared to NDa and SFa could thus
also be a finite Reynolds number effect,
since the adjustable helicity forcing consistently produced
simulations with lower Reynolds numbers for a given viscosity.

The turbulence is generated on a discrete Cartesian lattice,
so most points do not have integer values of $k$. 
Instead, a
shell-average of points with wavenumbers $n-0.5\leq k < n+0.5$,
where $n$ is a positive integer,
 is used when calculating spectral quantities.
This means that sometimes the density of states in a particular shell
will be higher or lower than the continuum limit of $4\pi k^2$,
causing bumps to appear in the spectra.
To counteract this, we have taken the spectral energy densities in the $n^{th}$ shell $S_n$ to be
$E^n_{u,b}=\frac{4\pi n^2}{M_n}\sum_{\vk\in S_n} E_{u,b}(\vk)$
where $M_n$ is the number of wavevectors in the shell \cite{Stepanov2014}.
This produces a smoother spectrum.

The magnetic and kinetic helicity generally remained negligible (see Table \ref{tab:param}),
however the cross helicity in some simulations was prone to
large fluctuations, yielding time-averaged values of 
$\rho_c$ as large as 0.22 (Run NDc).
We have seen that $\rho_c(k)$ is peaked at the forcing scales
(as shown in Fig. \ref{fig:ch_spectra})
and therefore we conclude that it has been injected by the forcing function.
Interestingly, 
there also tends to be a build-up in the normalised cross helicity at small scales
with the opposite sign to the build-up at the large scales.

Due to its effect on the scaling exponents of the energy spectra in 
MHD in the presence of a magnetic guide field \cite{Perez2009},
the behavior of $\sigma_c(k)$ has been studied extensively in solar wind data 
\cite{Podesta2010}, numerical simulations \cite{Beresnyak2010} and theoretically 
\cite{Podesta2011}. 
The observational studies found 
$|\sigma_c(k)|$ to be constant in the inertial subrange provided 
that the average value was large, 
while states with small mean cross helicity suffered from uncertainties in 
the measurements. 

Here, for steady states with low but non-negligible $\rho_c(k)$ and $\sigma_c(k)$,
we consistently find a tendency to compensate the force-induced 
alignment between $\vu$ and $\vb$ at small $k$ (large scales) mostly at large $k$ (small scales).
The relative cross helicity at a given scale is related to the scale-dependent 
alignment angle \cite{Boldyrev2006, Podesta2010}. However, provided the alignment angle is small, 
its scale dependence does not enforce a scale dependence of the 
relative cross helicity \cite{Podesta2010}. 
In view of the aforementioned results on the scale independence of 
$\sigma_c(k)$ for high-cross-helicity states in unbalanced MHD 
with a guide field, the approximately linear scaling observed here may 
point to a more complicated situation for low-cross-helicity states,
which merits further investigation in its own right. 
Finally, we do not find a measureable effect of nonzero cross helicity on the 
scaling of the energy spectra in our data. 

\subsection{Energy and dissipation ratios}
  \begin{figure}
    \centering
    \includegraphics[width=0.7\linewidth]{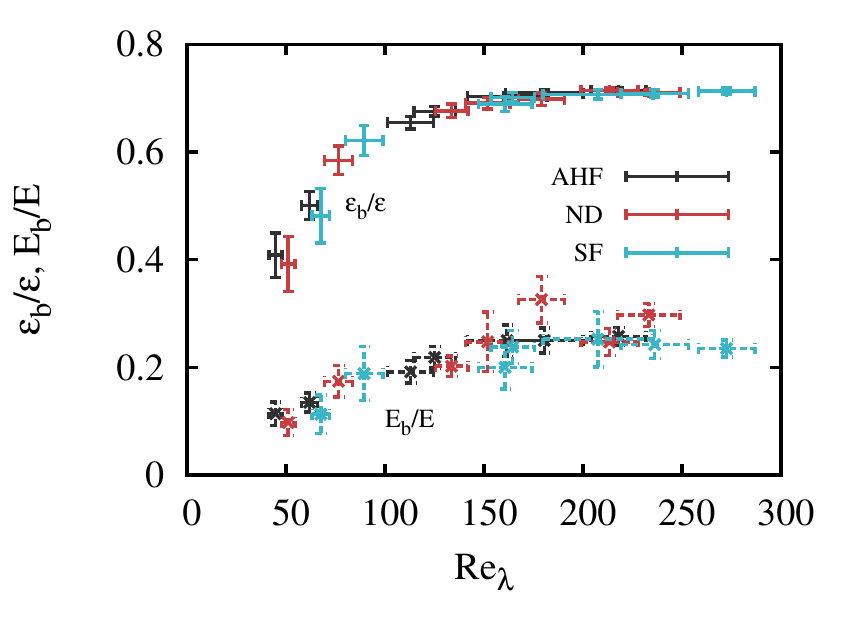}
    \caption{Fraction of magnetic dissipation $\epsilon_b/\epsilon$ (plusses) and magnetic energy $E_b/E$ (crosses)
            as a function of Taylor Reynolds number. 
            The error bars are the standard deviation. (Colour online.)}
    \label{fig:ratios}
  \end{figure}
In our tests, only the velocity field was forced and so the magnetic field
was sustained through the transfer of kinetic to magnetic energy,
i.e., dynamo action. 
It is useful to know how efficient the nonlinear dynamo is
at sustaining its magnetic field, which had initial conditions
such that it was in equipartition with
the velocity field at $t=0$.
Figure \ref{fig:ratios} shows the ratios $E_b/E$ and $\epsilon_b/\epsilon$
as a function of the Taylor-scale Reynolds number.
The Taylor-scale Reynolds number is used instead of the integral-scale
Reynolds number because we are interested in comparing the
effects of the forces at smaller scales than the forcing range.
The measurements of $E_b/E$ and $\epsilon_b/\epsilon$ follow a clear trend 
regardless of the way in which the kinetic energy was injected.
In particular, the magnetic dissipation fraction asymptotes quickly to 
$\epsilon_b/\epsilon\simeq0.71$.
This is in agreement with other results for
nonhelical simulations with unity magnetic Prandtl number \cite{Brandenburg2014,Haugen2003,Linkmann2017}.
The magnetic energy fraction displays slightly more erratic behavior,
particularly in run NDc.
The scatter is expected because the energy is dominated by the 
more volatile forcing scales,
while the dissipation takes place at small scales.
We conclude that the energy transfer and dissipation 
produced by each type of force are consistent.
The efficiency of the nonhelical nonlinear dynamo is independent of
the implementation of the large-scale 
mechanical forcing.

\subsection{Injection of ideal invariants}
\begin{figure}
  \begin{subfigure}{0.45\linewidth}
    \centering
    \includegraphics[width=0.9\linewidth]{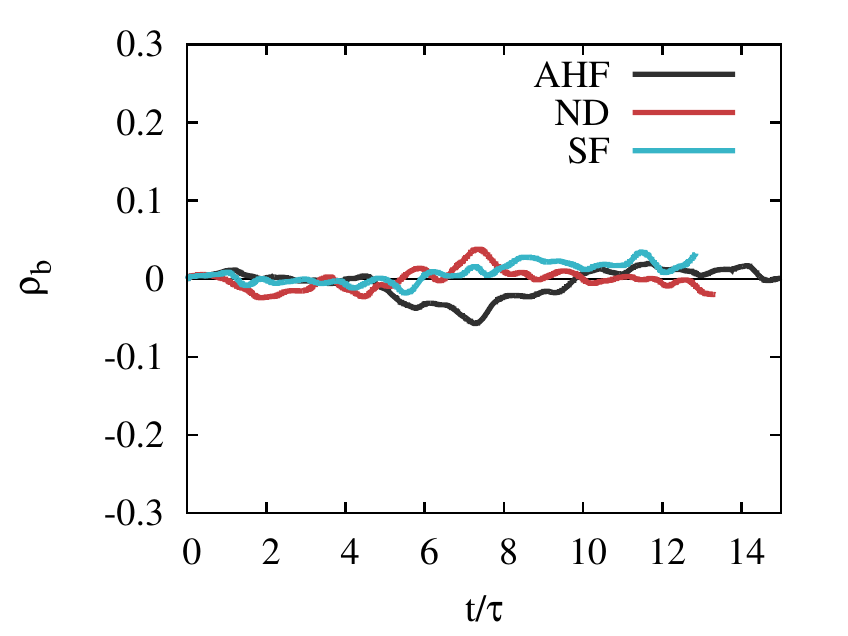}
    \caption{}
  \end{subfigure}
  \begin{subfigure}{0.45\linewidth}
    \centering
    \includegraphics[width=0.9\linewidth]{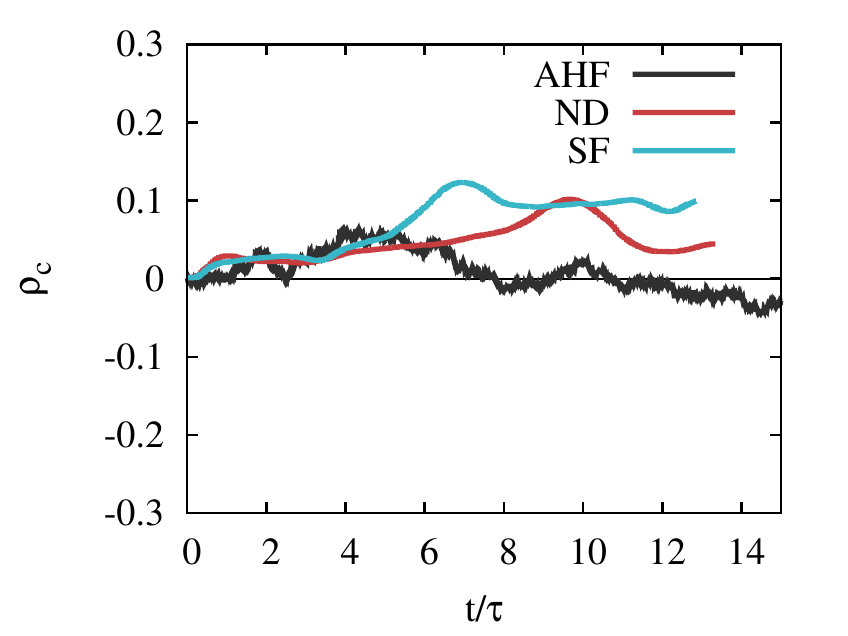}
    \caption{}
  \end{subfigure}
  \caption{Evolution of (a) relative magnetic helicity and (b) relative cross helicity
          for runs AHFa, NDa and SFa. 
           $\tau$ is the time-averaged large eddy turnover time.
           Note that the y-axis extends to $\pm0.3$ but the maximum possible values are 1. (Colour online.)
  \label{fig:hel}}
\end{figure}

The total energy, magnetic helicity and cross helicity are conserved in the 
ideal (non-dissipative) limit. It is therefore desirable 
for the helicities to remain approximately constant during a statistically
steady state at high Re and Rm. 
The total energy in the system fluctuates around a constant value
as energy is injected and dissipated and we expect the same from the other two ideal invariants.
The initial conditions in our simulations have zero magnetic and cross helicity.
The time evolution of relative magnetic helicity and cross helicity is shown in Fig. \ref{fig:hel}.
The mean relative magnetic helicity remains 
within one standard deviation of zero in all cases,
irrespective of the chosen forcing method.
This could be expected since the magnetic field is not directly forced
and should therefore be less susceptible to large variations.

The relative cross helicity, on the other hand, has the tendency to deviate from zero in some cases,
with large fluctuations lasting for long times.
This is particularly prevalent in the ND and SF runs,
with fluctuations up to $\rho_c\simeq0.3$ at times.
Since ND feeds the velocity field back into itself, small fluctuations in cross helicity
could be amplified, leading to a runaway effect at large scales.
In Fig. \ref{fig:ch_spectra} we saw that the relative cross helicity
is peaked at the forcing scales,
so it is clear that the growth of cross helicity is connected to the forcing. 
To guarantee negligible injection of cross helicity,
the alignment between $\vf$ and $\vb$ 
(and when a magnetic force $\vfr_\vbr$ is present, the alignment between $\vf_\vbr$ and $\vu$)
should remain negligible.
The unusually large fraction of magnetic energy in run NDc,
as seen in Fig. \ref{fig:ratios}, could be connected
to the presence of high cross helicity.
We will come back to this point in the next section.

The influence of intermediate values of cross helicity is not well understood,
although it is known that systems with nonzero cross helicity
can tend towards an Alfv\'enic state in which the cross helicity is maximal \cite{Grappin1983}.
One of the findings of this study that has not been anticipated in the literature
is that helicity can unexpectedly enter the system through certain forcing functions.
Thus it is of practical importance to monitor and control its injection.

\subsection{Comparison of repeated simulations}
Generally when using DNS we make the assumption of ergodicity,
.~e., we assume that in stationary turbulence the time-averaged values from one simulation
are equivalent to ensemble-averaged values
where the ensemble consists of many simulations.
Thus for any set of parameters, usually only one simulation is performed
and statistics are obtained by averaging over snapshots in time
once the system has reached a steady state.
This is the approach we took in the preceeding section.
However, we found significant variations in cross helicity
over time which prompted us to question how valid the ergodic principle
is in situations in which the injection of helicities 
cannot be controlled.
Furthermore, one might expect that the fluctuations of the ideal invariants would decrease as we increase the
Reynolds number, since we are increasing the number of interactions at each length scale,
but this does not seem to be the case in our tests.
Large fluctuations in cross helicity alter the behavior of a flow,
for example in Run NDc, which had an average relative cross helicity $\rho_c=0.22$
and a larger than expected energy fraction $E_b/E$.

\begin{figure}
    \centering
    \includegraphics[width=0.7\linewidth]{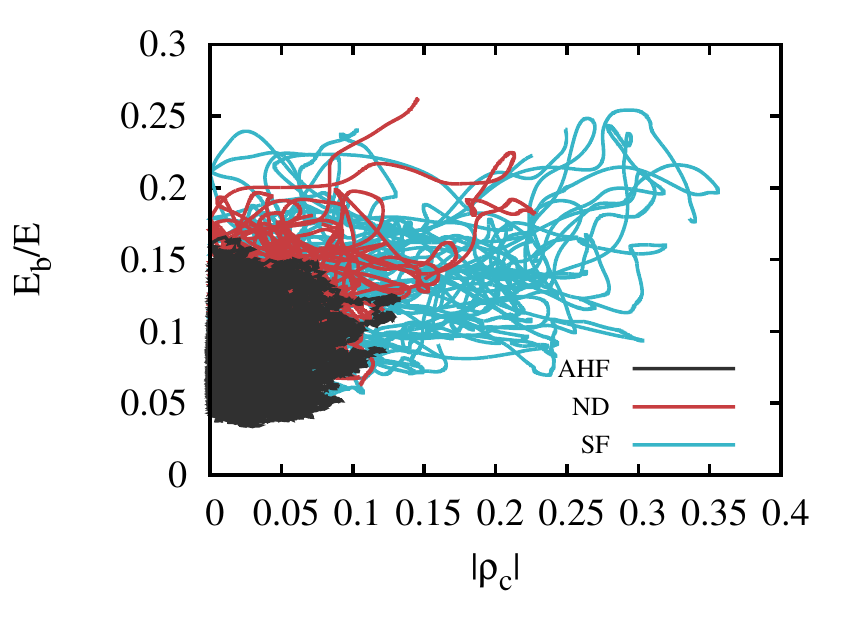}
    \caption{Magnitude of relative cross helicity versus the fraction of magnetic energy $E_b/E$
            at each point in time of the steady-state ensembles with $\nu=0.008$. (Colour online.)}
    \label{fig:ch_inf}
  \end{figure}

To test the variability of the cross helicity
and its effect on the distribution of energy between the two fields,
 we ran an ensemble of 20 simulations for each forcing type
on a $128^3$ grid with viscosity $\nu=0.008$.
The time-averaged Reynolds numbers and relative cross helicity $\rho_c$
are shown in Tab. \ref{tab:128}.
In all cases, the time-averaged Reynolds numbers stayed in a close range.
The relative cross helicity, however, was less consistent,
particularly in the SF ensembles. 
We plotted the time evolution of the magnetic energy fraction against the relative cross helicity
for all our simulations to explore the possibility of a connection
between the two quantities (Fig. \ref{fig:ch_inf}). 
We see further evidence of the variability of cross helicity in the SF cases
and also what seems to be a tendency for the magnetic energy fraction to increase 
as the magnitude of cross helicity increases.
This observation for fairly low levels of cross helicity is reminiscent of 
the behavior of the Archontis dynamo, which produces high levels of cross helicity
and saturates with the velocity and magnetic fields approximately in equipartition \cite{Archontis2007}.
Rapid variations in cross helicity have also been observed
in MHD shell models subject to a constant mechanical force \cite{Frick2000}.
Since the SF is static and the AHF is randomised at every time step,
the behavior we see also echoes that of Ref. \cite{Dallas2015}, who explored the effect of 
the correlation time of a kinetic and magnetic force
 on the helicities. Their results showed that the 
less often the force was randomised, the more the cross helicity was likely to grow.
We see a similar
effect here, although we only forced the velocity field.
\begin{table}[]
\centering
\begin{tabular}{ccccc}
Type & Re      & Re$_\lambda$ & $|\rho_c|$      & \# \\ \hline
AHF  & $80-87$   & $44-46$        & $7.41\times10^{-6}-0.0256$ & 20 \\
ND   & $100-111$ & $52-57$        & $0.000686-0.0561$ & 20 \\
SF   & $132-141$ & $66-70$        & $0.00111-0.193$ & 20 
\end{tabular}
\caption{The range of time-averaged quantities (integral-scale Reynolds number Re, Taylor-scale Reynolds number Re$_\lambda$
and relative cross helicity magnitude $|\rho_c|$) from ensembles of each of the three forcing types with $\nu=0.008$ 
on $128^3$ grid points.
}
\label{tab:128}
\end{table}

The results of this section suggest that the ergodic principle does not hold well
when the fluid is subject to the static sinusoidal forcing, 
as the relative cross helicity can vary significantly from one run to another,
which affects aspects of the system such as the energy distribution.
This is also likely the case with ND to a lesser degree.
The concept of nonuniversality is of interest here,
since two simulations with large average values of cross helicity
of opposite sign
would surely not behave in the same way as a system with zero cross helicity,
despite the ensemble-averaged values being small.
So perhaps the solution is simply to monitor the ideal invariants carefully
and be wary of large variations.
Nevertheless, the AHF reliably maintains small values of cross helicity
and so these extra considerations are not required.

\section{Conclusions\label{C}}
In this paper we explored the similarities and differences of three 
different types of mechanical 
forcing function in homogeneous, incompressible magnetohydrodynamic simulations
without a mean magnetic field.
In particular, we looked at negative damping, a random adjustable helicity forcing
in which the kinetic helicity input was set to zero,
and a nonhelical deterministic sinusoidal forcing.
From a practical point of view,
the AHF was least effective at reaching large Reynolds numbers
at a given resolution 
but most effective at maintaining small values of cross helicity.
It also produced slightly different energy spectra compared to the other two forcing types.
We found that all three forces produce a steady state in a similar amount of
simulation time, with reasonable agreement of dynamo efficiency,
interpreted via the energy and dissipation fractions.

We considered the fluctuations of energy, relative magnetic helicity and relative cross helicity 
over time since these three quantities are the ideal invariants in MHD.
The magnetic helicity had only very small fluctuations in all three cases,
whereas the cross helicity was more erratic. 
In some simulations - particularly the ND and SF types -
the cross helicity had large, long-term deviations from zero,
although the deviations were not large enough to cause the system to 
become fully Alfv\'enic.
However, it led us to question the validity of ergodicity when using forcing functions
which are prone to causing build-ups in cross helicity,
since large variations of cross helicity influence the development of the flow.
We found that there may be a tendency for the magnetic energy fraction to increase as 
the relative cross helicity increases, but the cross helicity fluctuations were not 
large enough to be able to say this definitively.
It is important to make sure that the rate of injection of cross helicity
is small.
Concerning the scale-by-scale behavior of 
the cross helicity measured through the relative 
cross-helicity spectrum, a response to large-scale 
cross helicity injection in the form of a mostly 
small-scale compensation effect was observed. Unlike 
high-cross-helicity states such as those observed in the solar wind 
\cite{Podesta2010}, the relative cross-helicity
spectra measured for our low-cross-helicity states 
are not scale-independent in the inertial subrange. 
This difference and the possible effect of a guide field
on cross helicity dynamics perhaps merits its own systematic 
investigation.

In summary, the present
analysis has highlighted some of the subtle problems with the 
control of ideal invariants in DNS of mechanically forced MHD turbulence.
Future work could involve carrying out more simulations with higher Reynolds numbers
to further assess the effect on fluctuations of the ideal invariants.
The AHF simulations had the best control of cross helicity injection,
presumably due to the stochastic nature of the force.
Adding a random phase to the SF function might therefore help to
minimize the cross helicity input.
It would also be interesting to study flows that are 
forced magnetically, with or without mechanical forcing,
particularly in light of the results on self-organization in
Ref.~\cite{Dallas2015}.

While we cannot make any absolute statements about the equivalence of all forcing functions
at large Reynolds numbers, we have at least confirmed that three different implementations,
typical of the kind generally used in MHD simulations,
produce flows with similar characteristics,
albeit with fairly significant deviations at the forcing scales. 
Forcing functions that do not control the injection of helicities
should be monitored carefully.
However, in the case of kinetic-only forcingi, upon which we have focused,
discrepancies introduced by different forcing functions have not been too large.
Thus, provided that the level of ideal invariants is maintained, it seems safe to rely on
the hypothesis that the small-scale behavior of a system is independent 
of how it is forced, at least in the case of 
mechanically-forced, homogeneous, nonhelical magnetohydrodynamics.\\


\begin{acknowledgments}
This work has made use of the resources provided by ARCHER \cite{ARCHER}, 
made available through the Edinburgh Compute and Data Facility 
(ECDF, \cite{ecdf}) and the Director's Time.
A.B.~acknowledges funding from the Science and Technology Facilities Council
and M.E.M.~from the Engineering and Physical Sciences Research Council (EP/M506515/1).
M.L.~thanks Luca Biferale for financial support of this project
through the European Union's Seventh Framework Programme (FP7/2007-2013)
under Grant Agreement No 339032.
We thank the anonymous referees for their useful suggestions.
The data supporting this publication are publicly available at the University of Edinburgh \cite{datashare}.
\end{acknowledgments}

\bibliographystyle{unsrt}
\bibliography{force}

\begin{thebibliography}{10}

\bibitem{Biskamp2003}
D.~Biskamp.
\newblock {\em Magnetohydrodynamic Turbulence}.
\newblock Cambridge University Press, Cambridge, UK, 2003.

\bibitem{Davidson2001}
P.~A. Davidson.
\newblock {\em An Introduction to Magnetohydrodynamics}.
\newblock Cambridge University Press, Cambridge, UK, 2001.

\bibitem{Frisch1995}
U.~Frisch.
\newblock {\em Turbulence: The Legacy of A. N. Kolmogorov}.
\newblock Cambridge University Press, Cambridge, UK, 1995.

\bibitem{Verma2004}
M.~K. Verma.
\newblock {Statistical theory of magnetohydrodynamic turbulence: Recent
  results}.
\newblock {\em Phys. Rep.}, 401(5-6):229--380, 2004.

\bibitem{Eswaran1988}
V.~Eswaran and S.~B. Pope.
\newblock An examination of forcing in direct numerical simulations of
  turbulence.
\newblock {\em Comput. Fluids}, 16(3):257--278, 1988.

\bibitem{Alvelius1999}
K.~Alvelius.
\newblock Random forcing of three-dimensional homogeneous turbulence.
\newblock {\em Physics of Fluids}, 11(7):1880--1889, 1999.

\bibitem{Zeren2010}
Z.~Zeren and B.~B\'edat.
\newblock Spectral and physical forcing of turbulence.
\newblock In {\em Progress in Turbulence III: Proceedings of the iTi Conference
  in Turbulence 2008}, pages 9--12. Springer Berlin Heidelberg, 2010.

\bibitem{McComb2014}
W.~D. McComb.
\newblock {\em Homogeneous, Isotropic Turbulence: Phenomenology,
  Renormalization and Statistical Closures}.
\newblock Oxford Science Publications, Oxford, UK, 2014.

\bibitem{Batchelor1953}
G.~K. Batchelor.
\newblock {\em The Theory of Homogeneous Turbulence}.
\newblock Cambridge University Press, Cambridge, UK, 1953.

\bibitem{Pope2000}
S.~B. Pope.
\newblock {\em Turbulent Flows}.
\newblock Cambridge University Press, Cambridge, UK, 2000.

\bibitem{Leorat1975}
J.~L\'eorat, U.~Frisch, and A.~Pouquet.
\newblock Helical magnetohydrodynamic turbulence and the nonlinear dynamo
  problem.
\newblock {\em Annals of the New York Academy of Sciences}, 257(1):173--176,
  1975.

\bibitem{Pouquet1976}
A.~Pouquet, U.~Frisch, and J.~L\'{e}orat.
\newblock Strong {MHD} helical turbulence and the nonlinear dynamo effect.
\newblock {\em J. Fluid Mech.}, 77:321--354, 1976.

\bibitem{Alexakis2005a}
A.~Alexakis, P.~D. Mininni, and A.~Pouquet.
\newblock On the inverse cascade of magnetic helicity.
\newblock {\em Astrophys. J.}, 640:335--343, 2006.

\bibitem{Alexakis2007}
A.~Alexakis, P.~D. Mininni, and A.~Pouquet.
\newblock Turbulent cascades, transfer, and scale interactions in
  magnetohydrodynamics.
\newblock {\em New Journal of Physics}, 9:298, 2007.

\bibitem{Moffat1969}
H.~K. Moffat.
\newblock The degree of knottedness of tangled vortex lines.
\newblock {\em J. Fluid Mech.}, 35:117--129, 1969.

\bibitem{Linkmann2015}
M.~F. Linkmann, A.~Berera, W.~D. McComb, and M.~E. McKay.
\newblock {Nonuniversality and Finite Dissipation in Decaying
  Magnetohydrodynamic Turbulence}.
\newblock {\em Phys. Rev. Lett.}, 114:235001, Jun 2015.

\bibitem{Linkmann2017}
M.~Linkmann, A.~Berera, and E.~E. Goldstraw.
\newblock Reynolds-number dependence of the dimensionless dissipation rate in
  homogeneous magnetohydrodynamic turbulence.
\newblock {\em Phys. Rev. E}, 95:013102, Jan 2017.

\bibitem{Dallas2013}
V.~Dallas and A.~Alexakis.
\newblock The signature of initial conditions of magnetohydrodynamic
  turbulence.
\newblock {\em Astrophys. J.}, 788(2):L36, 2014.

\bibitem{Dobrowolny1980}
M.~Dobrowolny, A.~Mangeney, and P.~Veltri.
\newblock {Fully Developed Anisotropic Hydromagnetic Turbulence in
  Interplanetary Space}.
\newblock {\em Phys. Rev. Lett.}, 45:144--147, Jul 1980.

\bibitem{Stribling1991}
T.~Stribling and W.~H. Matthaeus.
\newblock {Relaxation processes in a low-order three-dimensional
  magnetohydrodynamics model}.
\newblock {\em Physics of Fluids B: Plasma Physics}, 3(8):1848--1864, 1991.

\bibitem{Boldyrev2006}
S.~Boldyrev.
\newblock {Spectrum of Magnetohydrodynamic Turbulence}.
\newblock {\em Phys. Rev. Lett.}, 96:115002, 2006.

\bibitem{Dallas2015}
V.~Dallas and A.~Alexakis.
\newblock Self-organisation and non-linear dynamics in driven
  magnetohydrodynamic turbulent flows.
\newblock {\em Physics of Fluids}, 27(4):045105, 2015.

\bibitem{Archontis2007}
V.~Archontis, S.~B.~F. Dorch, and \AA. Nordlund.
\newblock Nonlinear {MHD} dynamo operating at equipartition.
\newblock {\em Astronomy and Astrophysics}, 472:715--726, 2007.

\bibitem{Dorch2004}
S.~B.~F. Dorch and V.~Archontis.
\newblock On the saturation of astrophysical dynamos: Numerical experiments
  with the no-cosines flow.
\newblock {\em Solar Physics}, 224(1):171--178, Oct 2004.

\bibitem{Gilbert2011}
A.~D. Gilbert, Y.~Ponty, and V.~Zheligovsky.
\newblock Dissipative structures in a nonlinear dynamo.
\newblock {\em Geophysical \& Astrophysical Fluid Dynamics}, 105(6):629--653,
  2011.

\bibitem{Cameron2006}
R.~Cameron and D.~Galloway.
\newblock High field strength modified {ABC} and rotor dynamos.
\newblock {\em Monthly Notices of the Royal Astronomical Society},
  367(3):1163--1169, 2006.

\bibitem{Yoffe2012}
S.~R. Yoffe.
\newblock {\em Investigation of the transfer and dissipation of energy in
  isotropic turbulence}.
\newblock PhD thesis, The University of Edinburgh, Scotland, 2012.

\bibitem{Linkmann2016}
M.~F. Linkmann.
\newblock {\em Self-organisation processes in (magneto)hydrodynamic
  turbulence}.
\newblock PhD thesis, The University of Edinburgh, Scotland, 2016.

\bibitem{Grappin1983}
R.~Grappin, J.~L\'eorat, and A.~Pouquet.
\newblock Dependence of {MHD} turbulence spectra on the velocity
  {field-magnetic} field correlation.
\newblock {\em Astronomy and Astrophysics}, 126(1):51--58, 1983.

\bibitem{Pouquet1988}
A.~Pouquet, P.~L. Sulem, and M.~Meneguzzi.
\newblock Influence of {velocity-magnetic} field correlations on decaying
  magnetohydrodynamic turbulence with neutral {X} points.
\newblock {\em The Physics of Fluids}, 31(9):2635--2643, 1988.

\bibitem{Machiels1997}
L.~Machiels.
\newblock Predictability of small-scale motion in isotropic fluid turbulence.
\newblock {\em Phys. Rev. Lett.}, 79:3411--3414, 1997.

\bibitem{Jimenez1993}
J.~Jim\'enez, A.~A. Wray, P.~G. Saffman, and R.~S. Rogallo.
\newblock The structure of intense vorticity in isotropic turbulence.
\newblock {\em Journal of Fluid Mechanics}, 255:65–90, 1993.

\bibitem{Kaneda2003}
Y.~Kaneda, T.~Ishihara, M.~Yokokawa, K.~Itakura, and A.~Uno.
\newblock Energy dissipation rate and energy spectrum in high resolution direct
  numerical simulations of turbulence in a periodic box.
\newblock {\em Physics of Fluids}, 15(2):L21--L24, 2003.

\bibitem{Kaneda2006}
Y.~Kaneda and T.~Ishihara.
\newblock High-resolution direct numerical simulation of turbulence.
\newblock {\em Journal of Turbulence}, 7:N20, 2006.

\bibitem{McComb2015}
W.~D. McComb, M.~F. Linkmann, A.~Berera, S.~R. Yoffe, and B.~Jankauskas.
\newblock Self-organization and transition to turbulence in isotropic fluid
  motion driven by negative damping at low wavenumbers.
\newblock {\em Journal of Physics A: Mathematical and Theoretical},
  48(25):25FT01, 2015.

\bibitem{Linkmann2015a}
M.~F. Linkmann and A.~Morozov.
\newblock Sudden relaminarization and lifetimes in forced isotropic turbulence.
\newblock {\em Phys. Rev. Lett.}, 115:134502, 2015.

\bibitem{Sahoo2011}
G.~Sahoo, P.~Perlekar, and R.~Pandit.
\newblock Systematics of the magnetic-{P}randtl-number dependence of
  homogeneous, isotropic magnetohydrodynamic turbulence.
\newblock {\em New Journal of Physics}, 13(1):013036, 2011.

\bibitem{Brandenburg2001}
A.~Brandenburg.
\newblock The inverse cascade and nonlinear alpha-effect in simulations of
  isotropic helical hydromagnetic turbulence.
\newblock {\em The Astrophysical Journal}, 550(2):824, 2001.

\bibitem{Malapaka2013}
S.~K. Malapaka and W.-C. M\"uller.
\newblock Large-scale magnetic structure formation in three-dimensional
  magnetohydrodynamic turbulence.
\newblock {\em The Astrophysical Journal}, 778(1):21, 2013.

\bibitem{Muller2012}
W.~C. M\"uller, S.~K. Malapaka, and A.~Busse.
\newblock Inverse cascade of magnetic helicity in magnetohydrodynamic
  turbulence.
\newblock {\em Phys. Rev. E.}, 85:015302, 2012.

\bibitem{Biferale2012}
L.~Biferale, S.~Musacchio, and F.~Toschi.
\newblock Inverse energy cascade in three-dimensional isotropic turbulence.
\newblock {\em Phys. Rev. Lett.}, 108:164501, Apr 2012.

\bibitem{Brandenburg2014}
A.~Brandenburg.
\newblock Magnetic {P}randtl number dependence of the kinetic-to-magnetic
  dissipation ratio.
\newblock {\em The Astrophysical Journal}, 791(1):12, 2014.

\bibitem{Galanti1992}
B.~Galanti, P.~L. Sulem, and A.~Pouquet.
\newblock Linear and non-linear dynamos associated with {ABC} flows.
\newblock {\em Geophysical \& Astrophysical Fluid Dynamics}, 66(1-4):183--208,
  1992.

\bibitem{Galloway2012}
D.~Galloway.
\newblock {ABC} flows then and now.
\newblock {\em Geophysical \& Astrophysical Fluid Dynamics}, 106(4-5):450--467,
  2012.

\bibitem{Mininni2007}
P.~D. Mininni.
\newblock Inverse cascades and $\ensuremath{\alpha}$ effect at a low magnetic
  {P}randtl number.
\newblock {\em Phys. Rev. E}, 76:026316, Aug 2007.

\bibitem{Childress1970}
S.~Childress.
\newblock New solutions of the kinematic dynamo problem.
\newblock {\em Journal of Mathematical Physics}, 11(10):3063--3076, 1970.

\bibitem{Galloway1984}
D.~Galloway and U.~Frisch.
\newblock A numerical investigation of magnetic field generation in a flow with
  chaotic streamlines.
\newblock {\em Geophysical \& Astrophysical Fluid Dynamics}, 29(1-4):13--18,
  1984.

\bibitem{Alexakis2013}
A.~Alexakis.
\newblock {Large-Scale Magnetic Fields in Magnetohydrodynamic Turbulence}.
\newblock {\em Phys. Rev. Lett.}, 110:084502, Feb 2013.

\bibitem{Stepanov2014}
R.~Stepanov, F.~Plunian, M.~Kessar, and G.~Balarac.
\newblock Systematic bias in the calculation of spectral density from a
  three-dimensional grid.
\newblock {\em Phys. Rev. E}, 90:053309, 2014.

\bibitem{Perez2009}
J.~C. Perez and S.~Boldyrev.
\newblock Role of cross-helicity in magnetohydrodynamic turbulence.
\newblock {\em Phys. Rev. Lett.}, 102:025003, 2009.

\bibitem{Podesta2010}
J.~J. Podesta and J.~E. Borovsky.
\newblock Scale invariance of normalized cross-helicity throughout the inertial
  range of solar wind turbulence.
\newblock {\em Physics of Plasmas}, 17(11):112905, 2010.

\bibitem{Beresnyak2010}
A.~Beresnyak and A.~Lazarian.
\newblock Scaling laws and diffuse locality of balanced and imbalanced
  magnetohydrodynamic turbulence.
\newblock {\em Astrophys. J. Lett.}, 722:L110, 2010.

\bibitem{Podesta2011}
J.~J. Podesta.
\newblock On the cross-helicity dependence of the energy spectrum in
  magnetohydrodynamic turbulence.
\newblock {\em Physics of Plasmas}, 18:012907, 2011.

\bibitem{Haugen2003}
N.~E.~L. Haugen, A.~Brandenburg, and W.~Dobler.
\newblock Is nonhelical hydromagnetic turbulence peaked at small scales?
\newblock {\em The Astrophysical Journal Letters}, 597(2):L141, 2003.

\bibitem{Frick2000}
{Frick, P.}, {Boffetta, G.}, {Giuliani, P.}, {Lozhkin, S.}, and {Sokoloff, D.}
\newblock Long-time behavior of mhd shell models.
\newblock {\em Europhys. Lett.}, 52(5):539--544, 2000.

\bibitem{ARCHER}
\url{http://www.archer.ac.uk/}.

\bibitem{ecdf}
\url{http://www.ecdf.ac.uk/}.

\bibitem{datashare}
\url{http://dx.doi.org/10.7488/ds/1999}.

\end{thebibliography}
\end{document}